\documentclass[aps,prd,reprint,onecolumn,showpacs,preprintnumbers, amsmath,amssymb,floatfix, nofootinbib,superscriptaddress]{revtex4-2}
\usepackage{graphicx}
\usepackage{dcolumn}
\usepackage{bm}
\usepackage{amsmath}

\usepackage[dvipsnames]{xcolor}

\definecolor{persianblue}{rgb}{0.11, 0.22, 0.73}

\newcommand*{\hc}{\mathrm{H.c.}}

\usepackage[
	colorlinks=true,
	citecolor=green!50!black,
        linkcolor=persianblue!70!black,
	urlcolor=green!50!black,
	hypertexnames=false]{hyperref}

\usepackage[mathlines]{lineno}
\usepackage{color}
\usepackage{xcolor}

\begin{document}
\title{Study of nucleon and $\Delta$ resonances from a systematic analysis of $K^\ast \Sigma$ photoproduction}
\author{Jun Shi}
\email{jun.shi@scnu.edu.cn}
\affiliation{State Key Laboratory of Nuclear Physics and Technology, Institute of Quantum Matter, South China Normal University, Guangzhou 510006, China}
\affiliation{Guangdong Basic Research Center of Excellence for Structure and Fundamental Interactions of Matter, Guangdong Provincial Key Laboratory of Nuclear Science, Guangzhou 510006, China}

\author{Bing-Song Zou}
\email{zoubs@mail.tsinghua.edu.cn}
\affiliation{Department of Physics, Tsinghua University, Beijing 100084, China}
\affiliation{CAS Key Laboratory of Theoretical Physics, Institute of Theoretical Physics,
Chinese Academy of Sciences, Beijing 100190, China}
\affiliation{School of Physics, University of Chinese Academy of Sciences (UCAS), Beijing 100049, China}
\affiliation{Southern Center for Nuclear-Science Theory (SCNT),
Institute of Modern Physics, Chinese Academy of Sciences, Huizhou 516000, China}

\begin{abstract}
A systematic analysis of the $K^\ast\Sigma$ photoproduction off proton is performed with all the available differential cross section data.
We carry out a strategy different from the previous studies of these reactions, where, instead of fixing the parameters of the added resonances from PDG or pentaquark models, we add the resonance with a specific $J^P$ and leave its parameters to be determined from the experimental data.
When adding only one resonance, the best result is to add a $N^\ast(3/2^-)$ around $2097$ MeV, which substantially reduces the $\chi^2$ per degree of freedom to $1.35$.
There are two best solutions when adding two resonances, one is to add one $N^\ast(3/2^-)$ and one $N^\ast(7/2^-)$, the other is to add one $N^\ast(3/2^-)$ and one $\Delta^\ast(7/2^-)$, leading the $\chi^2$ per degree of freedom to be $1.10$ and $1.09$, respectively.
The mass values of the $N^\ast(3/2^-)$ in the two resonance solutions are both near $2070$ MeV.
Our solutions indicate that the $3/2^-$ nucleon resonance around $2080$ MeV is strongly coupled with the $K^\ast\Sigma$ final state and support the molecular picture of $N(2080, 3/2^-)$.

\end{abstract}

\maketitle
\section{Introduction}

Baryon resonances have been widely studied since they can help to understand the inner structure of hadrons and reveal their underlying dynamics, as well as to explain the puzzle of missing nucleon resonances~\cite{Isgur:1977ef,Isgur:1978xj}.
The observations of the $P_c(4312)$, $P_c(4440)$, and $P_c(4457)$ states by LHCb experiments~\cite{LHCb:2015yax,LHCb:2019kea} have been the most convincing evidence for pentaquark states~\cite{Chen:2016qju,Guo:2017jvc,Zou:2021sha}.
These three narrow states can be explained as molecular bound states
with hidden charm of the $\bar{D}\Sigma_c$, $\bar{D}\Sigma_c^\ast$, and $\bar{D}^\ast\Sigma_c$ systems~\cite{Wu:2010jy,Wu:2010vk,Wang:2011rga,Wu:2019adv,Lin:2017mtz}, respectively.
The nucleon resonances $N(1875)3/2^-$ and $N(208)3/2^-$ which sit just below the $K\Sigma^\ast$ and $K^\ast\Sigma$ thresholds, are proposed to be the strange partners~\cite{He:2017aps,Lin:2018kcc,Zou:2018uji} of the $P_c$ molecular states.
In addition, some $P_c$ states are predicted as $\bar{D}^\ast\Sigma_c^\ast$ bound states from heavy quark spin symmetry~\cite{Xiao:2013yca,Liu:2019tjn,Du:2019pij}.
Their analogous strangeness partner is $N(2270)$ with $J^P=1/2^-$ or $3/2^-$ or $5/2^-$ as the $K^\ast\Sigma^\ast$ molecular states~\cite{Wu:2023ywu}.
The study of nucleon resonances can help to verify these model predictions and refine the baryon spectrum. 

The photoproduction off nucleon processes are generally used to study nucleon and $\Delta$ resonances.
Among them, the $K^\ast\Sigma$ final state reactions are of great interest~\cite{Capstick:1998uh} mainly because of two reasons.
One is that the $K^\ast\Sigma$ photoproduction
has a higher threshold, which is advantageous in exploring resonances around and above $2$ GeV.
The other is that, since the final state contains $s\bar{s}$, it can help to investigate nucleon resonances with hidden strangeness as potential pentaquark states.

Current experimental data include the differential cross sections from the CLAS Collaboration at the Thomas Jefferson National Accelerator Facility (Jefferson Lab)~\cite{CLAS:2007kab}(CLAS2007) and the CBELSA/TAPS Collaboration at the Electron Stretcher and Accelerator (ELSA)~\cite{CBELSATAPS:2008mpu}(CBELSA/TAPS2008) for the $\gamma p\to K^{\ast 0}\Sigma^+$ reaction, and the CLAS Collaboration at Jefferson Lab~\cite{CLAS:2013qgi}(CLAS2013) for the $\gamma p\to K^{\ast +}\Sigma^0$ reaction.
Theoretically, Zhao {\it{et al}}.~\cite{Zhao:2001jw} presented the first calculation of the $\gamma p\to K^{\ast}\Sigma$ reactions using a quark model and provided a description of collective resonance excitations.
Oh and Kim~\cite{Oh:2006in} studied the $\gamma p\to K^{\ast 0}\Sigma^+$ reaction with the preliminary CLAS Collaboration data~\cite{Hleiqawi:2005sz} focusing on the role of the $t$-channel $\kappa$ exchange.
Kim {\it{et al.}}~\cite{Kim:2012pz,Kim:2013czq} investigated the $\gamma p\to K^{\ast}\Sigma$ reactions, emphasizing the role of baryon resonances near the threshold where $N(2080)3/2^-$~\footnote{Note that before the 2012 review of PDG~\cite{ParticleDataGroup:2012pjm}, all the evidence for $3/2^-$ nucleon resonances above 1800 MeV was assigned under a two-star $N(2080)$. This resonance is not directly related to the $N(2080)3/2^-$ molecular state mentioned in the first paragraph. In our following text, the notation $N(2080)$ refers to the $K^\ast\Sigma$ molecular state.}, $N(2090)1/2^-$, $N(2190)7/2^-$, $N(2200)5/2^-$, $\Delta(2150)1/2^-$, $\Delta(2200)7/2^-$, and $\Delta(2390)7/2^+$ listed in the Particle Data Group (PDG) review of 2012~\cite{ParticleDataGroup:2012pjm} were added together, and they found that these resonance contributions play a minor role in producing the cross sections~\cite{CLAS:2007kab,CBELSATAPS:2008mpu,CLAS:2013qgi}.
Wang {\it{et al}}.~\cite{Wang:2017tpe} took into account a minimum number of additional resonance in the $s$ channel through checking the near-threshold four- or three-star resonances advocated in the 2016 PDG review~\cite{ParticleDataGroup:2016lqr}, and found that the four-star $\Delta(1905)5/2^+$ resonance is essential to describe the experimental data~\cite{CLAS:2007kab,CLAS:2013qgi}.
Ben {\it{et al}}.~\cite{Ben:2023uev} considered the contributions of the molecular states $N(2080)3/2^-$ and $N(2270)3/2^-$ coupled to $K^\ast\Sigma$ in $S$ wave and found that adding these two resonances is compatible with the differential cross sections~\cite{CLAS:2007kab,CLAS:2013qgi}, leading to a $\chi^2$ per data point as $1.648$.

The previous studies~\cite{Kim:2012pz,Kim:2013czq,Wang:2017tpe,Ben:2023uev} of $\gamma p\to K^{\ast}\Sigma$ reactions usually considered the added resonances with fixed parameter values from the PDG or the pentaquark model predictions.
In this work, we carry out a strategy different from the previous literature, in which we add the nucleon or $\Delta$ resonance with a specific quantum number $J^P$ and leave its properties, namely its mass, width, and coupling constants, as free parameters. In this analysis, we first add one resonance with specific $J^P$ and find the one with the most significant contribution to the $\gamma p\to K^\ast\Sigma$ reactions, then check the improvement when adding another resonance. With our solutions determined from the differential cross section data, we present the predictions of the polarization observables, which can be checked by future experimental data.

This article is organized as follows.
In Sec.~\ref{sec:theoretical_form}, we display the basic formalism of our theoretical evaluation.
Then we show our results and the corresponding discussions in Sec.~\ref{sec:result}. A brief summary is presented in Sec.~\ref{sec:summary}.

\section{theoretical formalism}\label{sec:theoretical_form}

\begin{figure}[htbp]
    \centering
    \includegraphics[width = 0.4\textwidth]{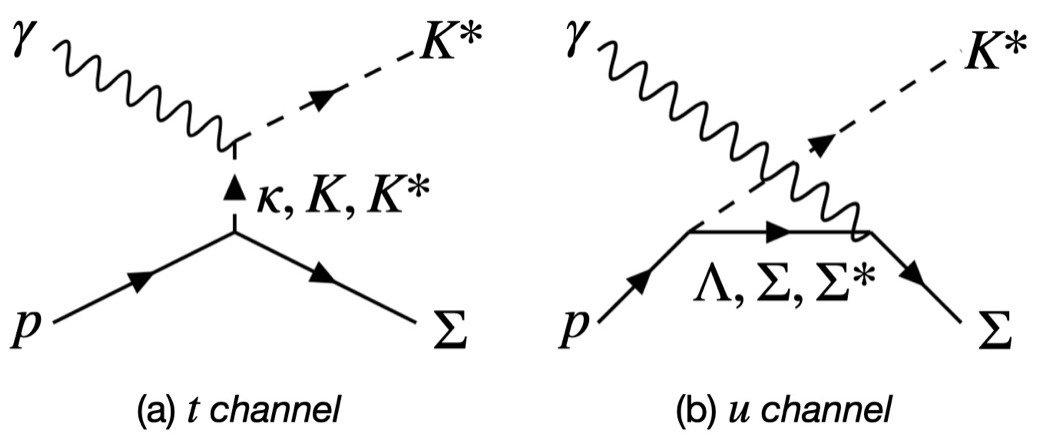}
    \includegraphics[width = 0.4\textwidth]{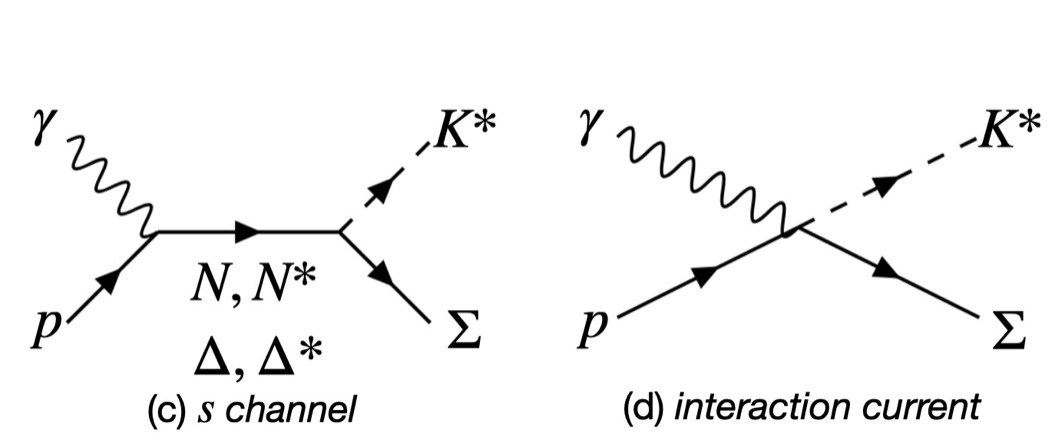}
    \caption{The Feynman diagrams of $\gamma p\to K^\ast \Sigma$. (a)–(c) t, u, s channels; (d) interaction current
    }
    \label{fig:feynman_diagram}
\end{figure}
The Feynman diagrams of the reactions $\gamma p\to K^{\ast 0}\Sigma^+$ and $\gamma p\to K^{\ast +}\Sigma^0$ are shown in Fig.~\ref{fig:feynman_diagram}.
Contributions from the $t$-channel $\kappa$, $K$, and $K^\ast$ exchanges, the $u$-channel $\Lambda$, $\Sigma$, and $\Sigma^\ast(1385)3/2^+$ exchanges, the $s$-channel nucleon and $\Delta(1232)1/2^+$ exchanges, as well as the interaction current where the photon interacts with the hadronic structure within the $K^\ast N\Sigma$ vertex are always included. Additional nucleon or $\Delta$ resonances with specific $J^P$ would also be added to the $s$ channel to further improve the description of the experimental data. 

We employ the effective Lagrangian approach at the tree-level Bonn approximation.
The effective Lagrangians for $\gamma p \to K^{\ast 0}\Sigma^+$ and $\gamma p \to K^{\ast+}\Sigma^0$ with respect to the electromagnetic interactions are expressed as~\cite{Oh:2006hm,Kim:2012pz}
\begin{eqnarray}
{\cal L}_{\gamma \kappa K^\ast} &=& g_{\gamma \kappa K^\ast}F^{\mu \nu}\bar{\kappa}K^\ast_{\mu \nu}, \\
{\cal L}_{\gamma K K^\ast} &=& g_{\gamma K K^\ast}\varepsilon^{\mu \nu \alpha \beta}\left(\partial_\mu A_\nu\right)\left(\partial_\alpha K^\ast_\beta\right)\overline{K}+H.c., \\
{\cal L}_{\gamma K^* K^* } &=& -ie_K^\ast A^\mu\left(K^{\ast-\nu}K^{\ast+}_{\mu\nu} - K^{\ast-}_{\mu\nu}K^{\ast+\nu}\right), \\
{\cal L}_{\gamma\Sigma \Sigma } &=& - \overline{\Sigma}\left[e_\Sigma\gamma^{\mu} 
- \frac{{e\kappa}_{\Sigma}}{2M_N}\sigma^{\mu\nu}\partial_\nu\right]A_\mu\Sigma,   \\
{\cal L}_{ \gamma\Sigma \Lambda} &=& \frac{e\mu_{\Sigma \Lambda}}{2M_N}\overline{\Lambda}\sigma^{\mu \nu}
\left(\partial_\nu A_\mu\right)\Sigma^0 + \hc,  \\
{\cal L}_{\gamma\Sigma^* \Sigma } &=& e\overline{\Sigma}^\ast_\mu\left[\frac{ig^{(1)}_{\gamma\Sigma^\ast \Sigma }}{2M_N}\gamma_\nu \gamma_5  + \frac{g^{(2)}_{\gamma\Sigma^\ast \Sigma }}{\left(2M_N\right)^2} \gamma_5\partial_\nu  \right]\Sigma F^{\mu \nu} + \hc,\\
{\cal L}_{\gamma NN} &=& - \overline{N} \left[ e_N \gamma^\mu
- \frac{ e{\kappa}_N} {2M_N}\sigma^{\mu \nu}\partial_\nu\right] A_\mu N, \\
{\cal L}_{\gamma\Delta N } &=& e\overline{\Delta}_{\mu}\left[\frac{ig^{(1)}_{\gamma\Delta N }}{2M_N}\gamma_\nu \gamma_5 
   + \frac{g^{(2)}_{\gamma\Delta N }}{\left(2M_N\right)^2} \gamma_5\partial_\nu \right] NF^{\mu \nu}
+ \hc,\label{eq:L_Deltagp}
\end{eqnarray}
where $e$ denotes the unit electric charge as $e=\sqrt{4\pi\alpha_{\text{EM}}}$ with $\alpha_{\text{EM}}=1/137.04$, and
$\kappa_B$ represents the anomalous magnetic moment of the baryon $B$ with values listed in Table~\ref{tab:fixed_bg}.
For the associated meson-baryon-baryon interactions, the effective Lagrangians are
\begin{eqnarray}
{\cal L}_{\kappa N\Sigma} &=& - g_{\kappa N\Sigma} \overline{\Sigma} \kappa N + \hc, \\
{\cal L}_{KN\Sigma} &=& -i g_{KN\Sigma}\bar{K}\overline{\Sigma}\gamma_{5} N + \hc, \label{eq:L_KNSigma}   \\
{\cal L}_{{K^\ast} N\Sigma} &=&  - g_{{K^\ast} N\Sigma} \overline{\Sigma} \left[\gamma^\mu-\frac{\kappa_{{K^\ast} N\Sigma}}{2M_N}\sigma^{\mu \nu}\partial_\nu \right]K^*_\mu N + \hc,\label{eq:L_KsNSigma}  \\
{\cal L}_{{K^\ast} N\Lambda} &=&  - g_{{K^\ast} N\Lambda} \overline{\Lambda} \left[\gamma^\mu-\frac{\kappa_{{K^\ast} N\Lambda}}{2M_N}\sigma^{\mu \nu}\partial_\nu \right]K^\ast_\mu N + \hc,   \nonumber  \\
{\cal L}_{K^\ast N \Sigma^\ast} &=& i\frac{g_{K^\ast N \Sigma^\ast}^{(1)}}{2M_{K^\ast}}\overline{N} \gamma_\nu \gamma_5\Sigma^\ast_\mu {K^\ast}^{\mu \nu} + \frac{g_{K^\ast N \Sigma^\ast}^{(2)}}{\left(2M_{K^\ast}\right)^2}\overline{N} \gamma_5 {\Sigma}^\ast_\mu{K^\ast}^{\mu \nu}\partial_\nu  - \frac{g_{K^\ast N \Sigma^\ast}^{(3)}}{\left(2M_{K^\ast}\right)^2}\overline{N} \gamma_5{\Sigma}^\ast_\mu\partial_\nu {K^{\ast\mu \nu}}  + \hc\label{eq:L_KsNSs} \\
{\cal L}_{K^\ast\Delta \Sigma} &=& - i\frac{g_{K^\ast\Delta \Sigma}^{(1)}}{2M_{K^\ast}}\overline{\Delta}_\mu \gamma_\nu \gamma_5 \Sigma{K^\ast}^{\mu \nu}  
- \frac{g_{K^\ast\Delta \Sigma}^{(2)}}{\left(2M_{K^\ast}\right)^2}\overline{\Delta}_\mu \gamma_5\partial_\nu\Sigma {K^\ast}^{\mu \nu}   - \frac{g_{K^\ast\Delta \Sigma}^{(3)}}{\left(2M_{K^\ast}\right)^2}\overline{\Delta}_\mu \gamma_5\Sigma\left(\partial_\nu {K^{\ast\mu \nu}}\right)  + \hc\label{eq:L_DeltaKS}
\end{eqnarray}
In the above Lagrangians, the isospin structure is always embedded.
For the coupling of isospin 1 and isospin $1/2$ particles,
given the $K N\Sigma$ coupling in Eq.~(\ref{eq:L_KNSigma}) as an example,
\begin{eqnarray}
K = \left(
\begin{array}{c} 
K^+ \\
K^0
\end{array} \right),~~~~
    \Sigma=\bm{\tau}\cdot\bm{\Sigma}=\left(
\begin{array}{cc} 
\Sigma^0 & \sqrt{2}\Sigma^+ \\
\sqrt{2}\Sigma^- &{\Sigma}^0
\end{array} \right),~~~~~
N=\left(
\begin{array}{c} 
p \\
n
\end{array} \right),
\end{eqnarray}
where $\bm{\tau}$ indicates the Pauli matrices.
The isospin structures of the $\Delta$
vertices in Eqs.~(\ref{eq:L_Deltagp}) and ~(\ref{eq:L_DeltaKS}) are, respectively, given as
follows~\cite{Kim:2012pz}:
\begin{eqnarray}
\overline\Delta I^0 N,~~~~ \overline\Delta
\bm{I}\cdot\bm{\Sigma}K^* ,
\end{eqnarray}
where $\bm I$ stands for the isospin transition ($3/2\to 1/2$) matrices  
\begin{equation}
\label{eq:iso_Delta}
I^- =
\frac{1}{\sqrt{6}}\left(\begin{array}{cc}
0&0\\
0&0\\
\sqrt{2}&0\\
0&\sqrt{6}
\end{array} \right),\,\,\,\,
I^0 =
\frac{1}{\sqrt{6}}\left(\begin{array}{cc}
0&0\\
2&0\\
0&2\\
0&0
\end{array} \right),\,\,\,\,
I^+ =
\frac{1}{\sqrt{6}}\left(\begin{array}{cc}
\sqrt{6}&0\\
0&\sqrt{2}\\
0&0\\
0&0
\end{array} \right).
\end{equation}

The effective Lagrangians for the $s$-channel resonance exchanges are~\cite{Kim:2012pz}
\begin{eqnarray}
\mathcal{L}_{\gamma  NR_{1/2^\pm}}  &=& 
\frac{ef_1}{2M_N} \overline{N} \Gamma^{(\mp)}\sigma_{\mu\nu}
\partial^\nu A^\mu R + \mathrm{H.c.}, \label{eq:LR1pgamma}                      
\\
\mathcal{L}_{\gamma  NR_{3/2^\pm}} &=& 
-ie \left[ \frac{f_1}{2M_N} \overline{N} \Gamma_\nu^{(\pm)} - 
           \frac{if_2}{(2M_N)^2} \partial_\nu \overline{N}
           \Gamma^{(\pm)} \right] 
F^{\mu\nu} R_\mu  + \mathrm{H.c.},                       
\\
\mathcal{L}_{\gamma  NR_{5/2^\pm}}  &=& 
e \left[ \frac{f_1}{(2M_N)^2} \overline{N} \Gamma_\nu^{(\mp)} -
         \frac{if_2}{(2M_N)^3} \partial_\nu \overline{N}
         \Gamma^{(\mp)}    \right] 
\partial^\alpha F^{\mu\nu} R_{\mu\alpha} + \mathrm{H.c.},
\\
\mathcal{L}_{\gamma  NR_{7/2^\pm}}&=& 
ie \left[ \frac{f_1}{(2M_N)^3} \overline{N} \Gamma_\nu^{(\pm)} -
          \frac{if_2}{(2M_N)^4} \partial_\nu \overline{N}
          \Gamma^{(\pm)} \right] 
\partial^\alpha \partial^\beta F^{\mu\nu} R_{\mu\alpha\beta} +
\mathrm{H.c.},   
\label{eq:R_EM}
\end{eqnarray}
\begin{eqnarray}
\label{eq:STRONG}
\mathcal{L}_{ K^* \Sigma R_{1/2^\pm}}
&=&-\frac{1}{2M_N} \overline R \left[
g_1 \biggl( \pm \frac{\Gamma_\mu^{(\mp)} \Sigma \partial^2}{M_R \mp M_N}
-i \Gamma^{(\mp)} \partial_\mu \biggr)
-g_2 \Gamma^{(\mp)} \sigma_{\mu\nu} \Sigma \partial^\nu 
\right] K^{*\mu} + \mathrm{H.c.},
\\
\mathcal{L}_{ K^* \Sigma R_{3/2^\pm}}                                    
&=& i\overline R_\mu \left[ 
\frac{g_1}{2M_N} \Gamma_\nu^{(\pm)}\Sigma  \mp 
\frac{ig_2}{(2M_N)^2}  \Gamma^{(\pm)}\partial_\nu \Sigma \pm 
\frac{ig_3}{(2M_N)^2}  \Gamma^{(\pm)}\Sigma \partial_\nu 
\right] K^{*\mu\nu} + \mathrm{H.c.}, 
\label{eq:NsKsR3}
\\
\mathcal{L}_{ K^* \Sigma R_{5/2^\pm}}                                     
&=& \overline R_{\mu\alpha} \left[ 
\frac{g_1}{(2M_N)^2} \Gamma_\nu^{(\mp)}\Sigma  \pm 
\frac{ig_2}{(2M_N)^3} \Gamma^{(\mp)}\partial_\nu \Sigma  \mp 
\frac{ig_3}{(2M_N)^3} \Gamma^{(\mp)} \Sigma \partial_\nu 
\right] \partial^\alpha K^{*\mu\nu} + \mathrm{H.c.},                                        
\\
\mathcal{L}_{ K^* \Sigma R_{7/2^\pm}} 
&=& -i\overline R_{\mu\alpha\beta} \left[ 
\frac{g_1}{(2M_N)^3} \Gamma_\nu^{(\pm)} \Sigma \mp 
\frac{ig_2}{(2M_N)^4} \Gamma^{(\pm)}\partial_\nu \Sigma  \pm 
\frac{ig_3}{(2M_N)^4} \Gamma^{(\pm)}\Sigma  \partial_\nu 
\right] \partial^\alpha \partial^\beta K^{*\mu\nu} + \mathrm{H.c.}, \label{eq:LR7mKs}
\end{eqnarray}
where $R_{J^P}$ stands for the nucleon or
$\Delta$ resonance with specific spin $J$ and parity $P$.
$\Gamma^{(\pm)}$ and $\Gamma_\nu^{(\pm)}$ in the above expressions are 
defined as
\begin{equation}
\label{eq:Gamma_mu}
\Gamma^{(\pm)} = \left(
\begin{array}{c} 
\gamma_5 \\ \mathbf{1}
\end{array} \right),
\qquad
\Gamma_\mu^{(\pm)} = \left(
\begin{array}{c}
\gamma_\mu \gamma_5 \\ \gamma_\mu 
\end{array} \right).
\end{equation}

As for the propagator with $q$ as the momentum of the exchanged particle, we use
\begin{equation}
    \frac{1}{q^2 - M_{\kappa/K}^2}
\end{equation}
for $\kappa$ and $K$ meson exchanges,
\begin{equation}
    \frac{-g_{\mu\nu}+q_\mu q_\nu/M_{K^\ast}}{q^2 - M_{K^\ast}^2}
\end{equation}
for $K^\ast$ meson exchange,
\begin{equation}
    \frac{\not\!{q} + M_R}{q^2 - M_{R}^2}
\end{equation}
for the spin-$1/2$ baryon exchange,
\begin{eqnarray}
\frac{\not\!{q} + M_R}{q^2 - M_{R}^2}
\left[ - g_{\mu\nu} +\frac{1}{3}\gamma_\mu\gamma_\nu +
\frac{1}{3M_R} (\gamma_\mu q_\nu - \gamma_\nu q_\mu )
 +\frac{2}{3M^2_R}q_\mu q_\nu \right]
 \end{eqnarray}
for the spin-$3/2$ baryon exchange,
\begin{eqnarray}
    \frac{\not\!{q} + M_R}{q^2 - M_{R}^2}S_{\alpha\beta\mu\nu}(q,M_R)
\end{eqnarray}
for the spin-$5/2$ baryon exchange, and
\begin{eqnarray}
    \frac{\not\!{q} + M_R}{q^2 - M_{R}^2}\Delta_{{\beta_1}{\beta_2}{\beta_3};{\alpha_1}{\alpha_2}{\alpha_3}}(q,M_R)
\end{eqnarray}
for the spin-$7/2$ baryon exchange, with
\begin{equation}
\tilde{g}_{\mu\nu} = g_{\mu\nu} - \frac{q_\mu q_\nu}{M_R^2}, 
~~~~
\tilde{\gamma}_\mu = \gamma_\mu - \frac{q_\mu}{M^2}\not\!{q},
\end{equation}
\begin{eqnarray}
S_{\alpha\beta\mu\nu}(q,M_R)=\frac{1}{2}( \tilde g_{{\alpha}{\mu}} \tilde g_{{\beta}{\nu}} 
+\tilde g_{{\alpha}{\nu}} \tilde g_{{\beta}{\mu}})
-\frac{1}{5} \tilde g_{{\alpha}{\beta}} \tilde g_{{\mu}{\nu}} 
-\frac{1}{10}( \tilde \gamma_{\alpha} \tilde \gamma_{\mu} \tilde
g_{{\beta}{\nu}}  
+\tilde \gamma_{\alpha} \tilde \gamma_{\nu} \tilde g_{{\beta}{\mu}}
+\tilde \gamma_{\beta} \tilde \gamma_{\mu} \tilde g_{{\alpha}{\nu}} 
+\tilde \gamma_{\beta} \tilde \gamma_{\nu} \tilde g_{{\alpha}{\mu}} ),
\end{eqnarray}
\begin{eqnarray}
\Delta_{{\beta_1}{\beta_2}{\beta_3};{\alpha_1}{\alpha_2}{\alpha_3}}
(q,M_R)=
\frac{1}{36}\sum_{P(\alpha),P(\beta)}
&&\left[-\tilde g_{{\beta_1}{\alpha_1}} \tilde g_{{\beta_2}{\alpha_2}} 
\tilde g_{{\beta_3}{\alpha_3}} +\frac{3}{7}\tilde g_{{\beta_1}{\alpha_1}}
\tilde g_{{\beta_2}{\beta_3}} \tilde g_{{\alpha_2}{\alpha_3}}\right.\\\nonumber
&&\left. +\frac{3}{7} \tilde \gamma_{\beta_1} \tilde \gamma_{\alpha_1} 
 \tilde g_{{\beta_2}{\alpha_2}} \tilde g_{{\beta_3}{\alpha_3}}
 -\frac{3}{35} \tilde \gamma_{\beta_1} \tilde \gamma_{\alpha_1} 
 \tilde g_{{\beta_2}{\beta_3}} \tilde g_{{\alpha_2}{\alpha_3}} \right],
\end{eqnarray}
where $P(\alpha)$ and $P(\beta)$ indicate the permutation over $\alpha_1 $, $\alpha_2 $, $\alpha_3 $ and $\beta_1 $, $\beta_2 $, $\beta_3 $, respectively.
The decay width is included by replacing $M$ with $M-i\Gamma/2$ for the $t$-channel $\kappa$, $K^\ast$, $u$-channel
$\Sigma^\ast(1385)$, and $s$-channel nucleon and $\Delta$ resonance exchanges in the propagator expressions.

An optional form factor is attached to each vertex to account for the composite structure of hadrons. For the $t$-channel meson exchange, we adopt the form factor expressed as
\begin{eqnarray}
    f_M(q^2, \Lambda) = \frac{\Lambda^2 - m_M^2}{\Lambda^2 - q^2};
\end{eqnarray}
while for the $u$- and $s$-channel baryon exchanges, the form factor takes the form
\begin{eqnarray}
    f_B(q^2, \Lambda) = \frac{\Lambda^4}{\Lambda^4 + (q^2 - m_B^2)^2},
\end{eqnarray}
where $q$ and $\Lambda$ represent the four-momentum and cutoff parameter of the exchanged particle, respectively.
Other cutoff parametrizations are also tested and do not affect our main results.

The total amplitude of $\gamma p\to K^{\ast}\Sigma$ reaction can be expressed as
\begin{equation}
    \mathcal{M} = \mathcal{M}_t + \mathcal{M}_u + \mathcal{M}_s + \mathcal{M}_{\text{int}}
=\varepsilon^{\ast}_{\lambda^\prime,\nu}\bar{u}^{s^\prime}_\Sigma\left[\mathcal{M}^{\mu\nu}_t + \mathcal{M}^{\mu\nu}_u + \mathcal{M}^{\mu\nu}_s + \mathcal{M}^{\mu\nu}_{\text{int}}\right]u^s_N\epsilon_{\lambda,\mu},
\end{equation}
where $u^s_N$ and $u^{s^\prime}_\Sigma$ represent the Dirac spinors of the nucleon and $\Sigma$ with spin indices $s$ and $s^\prime$, respectively, $\epsilon^\mu_{\lambda}$ and $\varepsilon^\mu_{\lambda^\prime}$ indicate the polarization vectors of the photon and $K^\ast$ with polarization indices $\lambda$ and $\lambda^\prime$ , separately. The polarization vectors $\epsilon^\mu_{\lambda}$ and $\varepsilon^\mu_{\lambda^\prime}$ are defined as
\begin{eqnarray}
\epsilon^\mu=\Big\{
\begin{array}{l}
\epsilon_\parallel=(0,1,0,0)\\
\epsilon_\perp=(0,0,1,0)
\end{array},\hspace{2cm}
\varepsilon^\mu=\Bigg\{
\begin{array}{l}
\varepsilon_1=(0,\cos\theta,0,-\sin\theta)\\
\varepsilon_2=(0,0,1,0)\\
\varepsilon_3=\frac{1}{M_{K^*}}(|\bm{k^\prime}|,E^\prime\sin\theta,
0,E^\prime\cos\theta)  
\end{array},
\end{eqnarray}
where $\theta$ indicates the scattering angle between the incoming photon and the outgoing $K^\ast$, and $k^\prime$ and $E^\prime$ represent the momentum and energy of $K^\ast$.
$\mathcal{M}_t$, $\mathcal{M}_u$, $\mathcal{M}_s$, and $\mathcal{M}_{\text{int}}$ represent the amplitudes of the $t-$, $u-$, and $s-$channel contributions and the interaction current.
Following Refs.~\cite{Haberzettl:1997jg,Haberzettl:2006bn,Huang:2011as,Wang:2017tpe}, the interaction current contribution can be approximated in a generalized contact current form
\begin{equation}
   \mathcal{M}^{\mu\nu}_{\text{int}} = C^\mu\Gamma^\nu_{K^\ast N \Sigma}(k^\prime) + M^{\mu\nu}_{KR}f_t, 
\end{equation}
where $\mu$, $\nu$ are Lorentz indices of the polarization vectors of the photon and $K^\ast$, $\Gamma^\nu_{K^\ast N \Sigma}(q)$ indicates the vertex function of $K^\ast N\Sigma$ coupling from Eq.~(\ref{eq:L_KsNSigma}) and is expressed as
\begin{equation}
 \Gamma^\nu_{K^\ast N \Sigma}(k^\prime) = - g_{{K^\ast} N\Sigma}  \left[\gamma^\nu-i\frac{\kappa_{{K^\ast} N\Sigma}}{2M_N}\sigma^{\nu \alpha} k^{\prime\alpha}\right].
\end{equation}
$C^\mu$ represents the auxiliary current formulated as
\begin{eqnarray}
    C^\mu = -Q_{K^\ast}\frac{f_t - \hat{F}}{t-k^{\prime 2}}\left(2p-k\right)^\mu - Q_N\frac{f_s - \hat{F}}{s-p^{ 2}}\left(2p+k\right)^\mu - Q_\Sigma\frac{f_u - \hat{F}}{u-p^{\prime 2}}\left(2p^\prime-k\right)^\mu - Q_N\frac{f_s - \hat{F}}{s-p^{ 2}}\left(2p+k\right)^\mu,
\end{eqnarray}
with
\begin{equation}
    \hat{F} = 1-\hat{h}(1-\delta_sf_s)(1-\delta_u f_u)(1-\delta_t f_t),\label{eq:F_hat}
\end{equation}
where $k,~p,~k^\prime$, and $p^\prime$ represent the momentum of the initial photon and proton and the outgoing $K^\ast$ and $\Sigma$, respectively, and $Q_K^{\ast}$, $Q_N$, and $Q_\Sigma$ are the electric charge of $K^\ast$, $ Q_N$ and $Q_\Sigma$. $\delta_{s,t,u}$ is an auxiliary constant that selects the channels that contributed to a given scattering reaction. Specifically, $\delta_{s,u} = 1$ and $\delta_{t} = 0$ for $\gamma p\to K^{\ast 0}\Sigma^+$ reaction, while $\delta_{s,t} = 1$ and $\delta_{u} = 0$ for the reaction $\gamma p\to K^{\ast +}\Sigma^0$.
$f_t$, $f_u$, and $f_s$ indicate the form factor attached to the amplitude of the $t$-channel $K^\ast$, $u$-channel $\Sigma$, and $s-$channel nucleon, respectively.
$M^{\mu\nu}_{\text{KR}}$ is the Kroll-Ruderman term expressed as
\begin{equation}
    M^{\mu\nu}_{\text{KR}} = ig_{K^\ast N\Sigma}\frac{\kappa_{K^\ast N\Sigma}}{2M_N}\sigma^{\mu\nu}Q_{K^\ast}.
\end{equation}
$M^{\mu\nu}_{\text{KR}}$ comes from the effective Lagrangian of $\gamma N\to K^\ast\Sigma$,
\begin{equation}
    \mathcal{L}_{\gamma N K^\ast \Sigma} = -ig_{K^\ast N\Sigma}\frac{\kappa_{K^\ast N\Sigma}}{2M_N}\bar{\Sigma}\sigma^{\mu\nu}A_\nu \hat{Q}_{K^\ast}K^\ast_\mu N + \mathrm{H.c.},
\end{equation}
which is obtained by the minimal gauge substitution $\partial_\mu\to D_\mu\equiv \partial_\mu - i\hat{Q}_{K^\ast}A_\mu$ with $\hat{Q}_{K^\ast}$ as the electric charge operator for the outgoing $K^\ast$ meson.

The differential cross section can be expressed as
\begin{equation}
    \frac{d\sigma}{d\cos{\theta}} = \frac{1}{4}\frac{1}{32\pi s}\frac{|\bold{k^\prime}|}{|\bold{k}|}\left|\mathcal{M}\right|^2,
\end{equation}
where $1/4$ indicates the average of the polarizations of the initial photon and proton, $\bold{k}$ and $\bold{k^\prime}$ represent the three-momentum of the incoming photon and the outgoing $K^\ast$, respectively, $s$ is the center-of-mass (c.m.) energy of the reaction.

\begin{table}[h]
\begin{tabular}{|c|c|c|c|c|c|c|c|}
\hline 
 & $g_{\gamma\kappa^{+}K^{\ast+}}$ & $g_{\gamma\kappa^{0}K^{\ast0}}$ & $g_{\gamma K^{+}K^{\ast+}}$ & $g_{\gamma K^{0}K^{\ast0}}$ & $\kappa_{\Sigma^{+}}$ & $\kappa_{\Sigma^{+}}$ & $\kappa_{\Sigma^{0}}$ \\
\hline 
Value & $0.119e$ & $-0.238 e$ & $0.413$ & $-0.631$ & $1.458$ & $-0.16$ & $0.65$ \\
\hline 
Note(Ref.) & \multicolumn{2}{c|}{vector-meson-dominance model\cite{Oh:2006hm,Black:2002ek}} & \multicolumn{2}{c|}{decay width~\cite{ParticleDataGroup:2024cfk}} & PDG~\cite{ParticleDataGroup:2024cfk} & PDG~\cite{ParticleDataGroup:2024cfk} & $(\kappa_{\Sigma^{+}}+\kappa_{\Sigma^{-}})/2$~\cite{Wang:2017tpe} \\
\hline
\end{tabular} 
\newline
\vspace*{0.2 cm}
\newline
\centering
\begin{tabular}{|c|c|c|c|c|c|c|c|c|}
\hline
 & $\kappa_{\Sigma\Lambda}$ & $\kappa_{p}$ & $g_{\gamma\Delta N}^{(1)}$ & $g_{\gamma\Delta N}^{(2)}$ & $g_{\kappa N\Sigma}$ & $g_{KN\Sigma}$ & $g_{K^{\ast}N\Lambda}$ & $\kappa_{K^{\ast}N\Lambda}$\\
\hline 
Value & $-1.61$ & $1.793$ & $-4.18$ & $4.32$ & $-5.32$ & $2.692$  & $-6.19$ & $2.77$\\
\hline 
Note(Ref.) & PDG~\cite{ParticleDataGroup:2024cfk} & PDG~\cite{ParticleDataGroup:2024cfk} & \multicolumn{2}{c|}{helicity amplitude~\cite{Oh:2007jd,ParticleDataGroup:2024cfk}}& soft-core model~\cite{Stoks:1999bz} & SU(3)~\cite{Ronchen:2012eg} & SU(3)\cite{Ronchen:2012eg} & SU(3)~\cite{Ronchen:2012eg}\\
\hline 
\end{tabular}
\caption{The values and references of the fixed parameters in the background channels. }\label{tab:fixed_bg}
\end{table}

Most of the coupling constants of the background channels are determined either from the SU(3) symmetry, the decay width of the exchanged particle from experiments, or phenomenological models and would be fixed in our analysis.
We list their values and sources in Table~\ref{tab:fixed_bg}.
For the coupling constants of the $K^\ast N\Sigma$ interaction as listed in Eq.~(\ref{eq:L_KsNSigma}), Ref.~\cite{Stoks:1999bz} presented two sets of values from the Nijmegen soft-core models denoted as NSC97a and NSC97f with the magnetic vector $F/(F+D)$ ratio $\alpha^m_V$
being $0.4447$ and $0.3647$, respectively, as
\begin{eqnarray}
    g_{K^\ast N \Sigma}&=&-2.46,\hspace{3cm} \kappa_{{K^\ast} N\Sigma}=-0.47\hspace{1cm} (\text{NSC97a})\\
    g_{K^\ast N \Sigma}&=&-3.52,\hspace{3cm} \kappa_{{K^\ast} N\Sigma}=-1.14\hspace{1cm} (\text{NSC97f}),
\end{eqnarray}
while the estimated values of these two couplings from SU(3) symmetry~\cite{Ronchen:2012eg} is
\begin{equation}
    g_{K^\ast N \Sigma}=-4.23,\hspace{3cm} \kappa_{{K^\ast} N\Sigma}=-2.34.
\end{equation}
Since the $K^\ast N\Sigma$ interaction involves in many channels such as the $t$-channel $K^\ast$ exchange, the $u$-channel $\Sigma$ exchange, and the $s$-channel $N$ exchange, as well as the interaction current, we choose $g_{K^\ast N \Sigma}$ and $\kappa_{{K^\ast} N\Sigma}$ to be fitting parameters with limits $-4.23\sim -2.46$ and $-2.34\sim -0.47$, respectively.
Other free parameters of the background contributions include~\footnote{For $g_{K^\ast N \Sigma^\ast}^{(1)}$ in Eq.~(\ref{eq:L_KsNSs}), Ref.~\cite{Kim:2012pz} takes the value 5.2 while Ref.~\cite{Wang:2017tpe} uses 15.2, where the comparison has been adjusted to be matchable according to different factors in the $K^\ast N \Sigma^\ast$ Lagrangians of the two references. We thus take this constant multiplying the coupling constant of the electromagnetic interaction as a free parameter.} $g_{K^\ast N \Sigma^\ast}^{(1)}g^{(1,2)}_{\gamma\Sigma^{\ast +} \Sigma^{+} }$, $g_{K^\ast N \Sigma^\ast}^{(1)}g^{(1,2)}_{\gamma\Sigma^{\ast 0} \Sigma^{0} }$, $g_{K^\ast\Delta \Sigma}^{(1)}$, and the cutoff parameter of each channel.
The terms with $g_{K^\ast N \Sigma^\ast}^{(2,3)}$, and $g_{K^\ast\Delta \Sigma}^{(2,3)}$ are not considered in the present work following Refs.~\cite{Kim:2012pz,Wang:2017tpe}.

\section{Results and Discussion}\label{sec:result}

The differential cross section data we use for the reaction $\gamma p\to K^{\ast 0}\Sigma^+$ are from CLAS2007~\cite{CLAS:2007kab} with 48 data points and CBELSA/TAPS2008~\cite{CBELSATAPS:2008mpu} with 24 data points, while those for the $\gamma p\to K^{\ast +}\Sigma^0$ reaction are from CLAS2013~\cite{CLAS:2013qgi} with 178 data points.
The fitting includes 15 free parameters of the background contributions, that is, seven coupling constants and eight cutoff parameters, together with the parameters of the additional added $s$-channel nucleon or $\Delta$ resonance. Adding one resonance would result in five (for $1/2^\pm$ resonance) or eight (for $3/2^\pm$, $5/2^\pm$ or $7/2^\pm$ resonance) additional fitting parameters, including the cutoff parameter, the mass, width and 2 or 4 independent coupling constants\footnote{Only the multiplications of the coupling constants from the electromagnetic interaction and the meson-baryon-meson interaction as listed in Eqs.~(\ref{eq:LR1pgamma})-(\ref{eq:LR7mKs}) are independent. Given the $3/2^\pm$ resonance as an example, the independent coupling constants are $f_1g_1$, $f_2/f_1$, $g_2/g_1$, and $g_3/g_1$.} of the resonance.

With only the background contributions, the $\chi^2$ per degree of freedom is $\chi^2/\text{ndf}=3.40$ by using the MINUIT algorithm~\cite{James:1975dr,iminuit}, which is far from adequate to describe the experimental data. 

\begin{figure}[htbp]
    \centering
    \includegraphics[width = 0.8\textwidth]{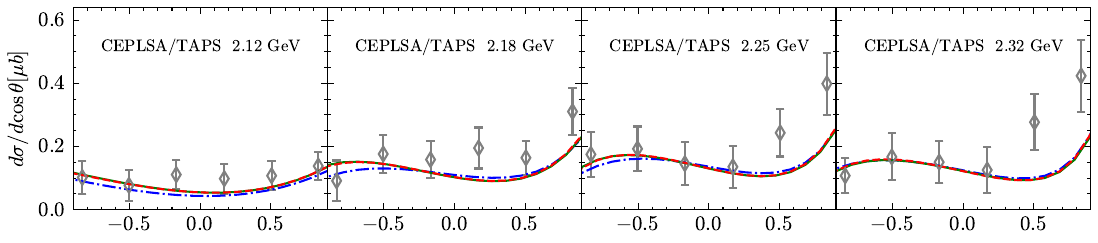}
    \includegraphics[width = 0.8\textwidth]{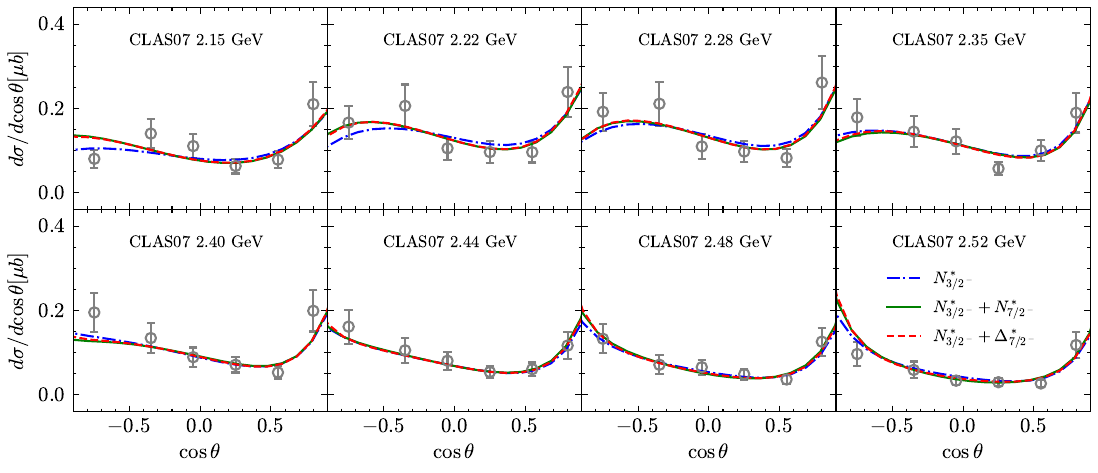}
    \caption{The differential cross sections for the reaction $\gamma p\to K^{\ast 0}\Sigma^+$, where the blue dash-dotted line, the green solid line and the red dashed line represent the solutions with adding one $N^\ast(3/2^-)$ resonance, $N^\ast(3/2^-)$ plus $N^\ast(7/2^-)$ resonances, and $N^\ast(3/2^-)$ plus $\Delta^\ast(7/2^-)$ resonances, respectively, comparing to the experimental data of CBELSA/TAPS2008~\cite{CBELSATAPS:2008mpu} (thin diamond, upper row) and CLAS2007~\cite{CLAS:2007kab} (circle, lower rows).
    }
    \label{fig:dcs_a}
\end{figure}

\begin{figure}[h]
    \centering
    \includegraphics[width = 0.8\textwidth]{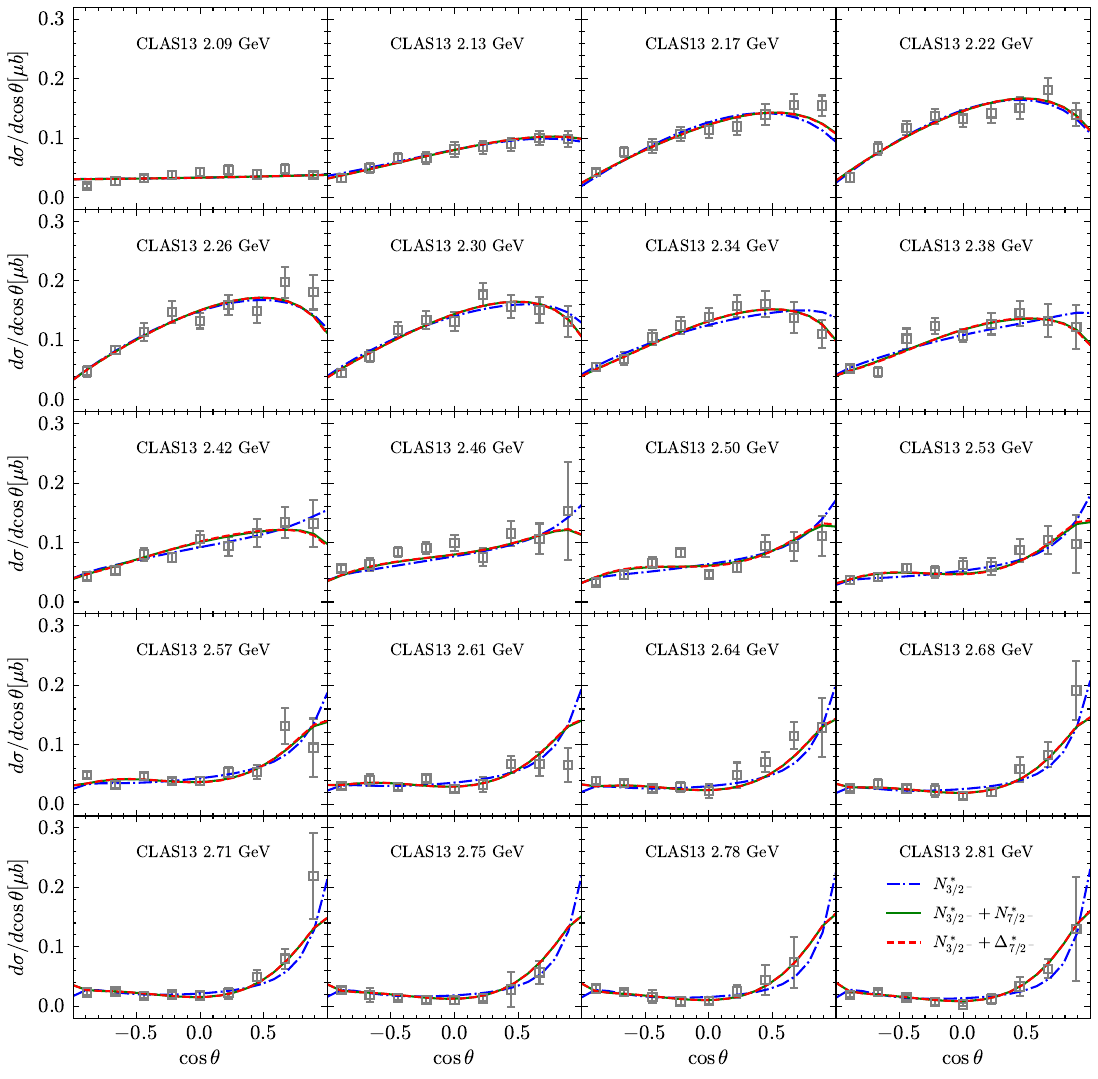}
    \caption{The differential cross sections for the reaction $\gamma p\to K^{\ast 0}\Sigma^+$, with the three solutions noted the same as Fig.~\ref{fig:dcs_a} comparing to the experimental data of CLAS13.
    }
    \label{fig:dcs_b}
\end{figure}

When adding only one nucleon or $\Delta$ resonance, the best solution is to add one ${3}/{2}^-$ nucleon resonance with its mass around $2.097$ GeV, and the corresponding $\chi^2/\text{ndf}=1.35$. Adding this $N^\ast({3}/{2}^-)$ substantially improves the description of the experimental data, which implies that the $N^\ast({3}/{2}^-)$ resonance at $2.097$ GeV is strongly coupled with the $K^\ast\Sigma$ final state.
In Figs.~\ref{fig:dcs_a} and \ref{fig:dcs_b}, we show the respective description of the experimental data for the reactions $\gamma p\to K^{\ast 0}\Sigma^+$ and $\gamma p\to K^{\ast +}\Sigma^0$ in dash-dotted lines when adding this resonance, where the energies noted in the plots represent the c.m. energy of the system.
We can see that the experimental data can be well described by this solution.
The fitted parameters of this solution are listed in Table~\ref{tab:solution_1rN3m}.
We list other results with $\chi^2/\text{ndf}$ less than $2.0$ in Table~\ref{tab:solution_1r_other} together with the mass and width of the added resonance.
Their possible corresponding PDG listed particles with ratings of establishment are also noted according to their mass values.

\begin{table}[h]
\begin{tabular}{|c|c|c|c|c|c|c|}
\hline 
  $g_{K^{\ast}N\Sigma}$ & $\kappa_{K^{\ast}N\Sigma}$ & $g^{(1)}_{K^{\ast}N\Sigma^{\ast}}g^{(1)}_{\gamma\Sigma^{\ast+}\Sigma^+}$ & $g^{(1)}_{K^{\ast}N\Sigma^{\ast}}g^{(2)}_{\gamma\Sigma^{\ast+}\Sigma^+}$ & $g^{(1)}_{K^{\ast}N\Sigma^{\ast}}g^{(1)}_{\gamma\Sigma^{\ast0}\Sigma^0}$ & $g^{(1)}_{K^{\ast}N\Sigma^{\ast}}g^{(2)}_{\gamma\Sigma^{\ast0}\Sigma^0}$ & $g^{(1)}_{K^\ast\Delta\Sigma}$ \tabularnewline
\hline 
$-2.46(1.41)$ &$-2.34(2.01)$ &$7.33(5.15)$ &$-0.003(6.601)$ &$-0.33(0.73)$ &$-8.13(6.08)$ & $40.00(9.05)$ \\
\hline
\end{tabular} 
\newline
\vspace*{0.2 cm}
\newline
\begin{tabular}{|c|c|c|c|c|c|}
\hline 
  $M_R$ (MeV) & $\Gamma_R$ (MeV) & $f_1g_1$ & $f_2/f_1$ & $g_2/g_1$ & $g_3/g_1$ \tabularnewline
\hline 
$2095.9(5.7)$& $66.3(10.0)$ & $-5.47(0.31)$ & $-0.17(0.072)$ & $-0.52(0.06)$ & $1.36(0.08)$ \\
\hline
\end{tabular}
\caption{Fitted parameters when adding one ${3/2}^-$ resonance with $\chi^2/\text{ndf} = 1.35$.}\label{tab:solution_1rN3m}
\end{table}

\begin{table}[h]
\begin{tabular}{|c|c|c|c|c|}
\hline 
& $\chi^2/\text{ndf}$ & $M_R$ (MeV) & $\Gamma_R$ (MeV) & PDG~\cite{ParticleDataGroup:2024cfk}\\
\hline
$N^\ast(1/2^-)$ & $1.77$ & $2072.1(9.8)$ & $10.4(5.46)$ & Four-star $N(1895)$\\
\hline
$N^\ast(5/2^-)$ & $1.67$ & $2085.9(21.9)$ & $317.5(77.8)$ & Three-star $N(2060)$\\
\hline
$\Delta^\ast(1/2^-)$ & $1.66$ & $2081.7(6.1)$ & $10.0(19.6)$ & Four-star $\Delta(1950)$ or one-star $\Delta(2150)$\\
\hline
$\Delta^\ast(3/2^+)$ & $1.81$ & $2122.5(50.1)$ & $400.0(240.9)$ & Three-star $\Delta(1920)$\\
\hline
$\Delta^\ast(3/2^-)$ & $1.44$ & $2092.0(11.7)$ & $69.8(11.5)$ & One-star $\Delta(1940)$\\
\hline
$\Delta^\ast(5/2^-)$ & $1.75$ & $2065.8(26.9)$ & $400.0(69.9)$ & Three-star $\Delta(1930)$\\
\hline
\end{tabular}
\caption{The mass and width of the added resonance $\chi^2/\text{ndf}$ lower than $2.0$ when adding only one resonance, together with the corresponding particles in the PDG~\cite{ParticleDataGroup:2024cfk} list with establishment status noted in one to four stars.}\label{tab:solution_1r_other}
\end{table}

When adding two additional nucleon or $\Delta$ resonances, there are two best results. One is to add one $N^\ast({3}/{2}^-)$ and one $N^\ast({7}/{2}^-)$, the other is to add one $N^\ast({3}/{2}^-)$ and one $\Delta^\ast({7}/{2}^-)$, with the resulting $\chi^2/\text{ndf}$ being $1.10$ and $1.09$, respectively. The corresponding parameters of these two solutions are listed in Tables~\ref{tab:solution_2rN3mN7m} and \ref{tab:solution_2rN3mD7m}.

Both solutions include a $N^\ast({3}/{2}^-)$ around $2070$ GeV, which again supports the strong coupling of the $N^\ast({3}/{2}^-)$ similar to the adding one additional resonance condition.
The mass of the $N^\ast({7/2}^-)$ resonance and $\Delta^\ast({7/2}^-)$ is $2433.3$ and $2437.1$ MeV, respectively.
They may correspond to the four-star $N(2190)$ and the three-star $\Delta(2200)$ listed in PDG~\cite{ParticleDataGroup:2024cfk}.
Adding other resonances would result in an increase of the $\chi^2$ by at least $30$ and are not considered as appropriate solutions.

\begin{table}[h]
\begin{tabular}{|c|c|c|c|c|c|c|}
\hline 
  $g_{K^{\ast}N\Sigma}$ & $\kappa_{K^{\ast}N\Sigma}$ & $g^{(1)}_{K^{\ast}N\Sigma^{\ast}}g^{(1)}_{\gamma\Sigma^{\ast+}\Sigma^+}$ & $g^{(1)}_{K^{\ast}N\Sigma^{\ast}}g^{(2)}_{\gamma\Sigma^{\ast+}\Sigma^+}$ & $g^{(1)}_{K^{\ast}N\Sigma^{\ast}}g^{(1)}_{\gamma\Sigma^{\ast0}\Sigma^0}$ & $g^{(1)}_{K^{\ast}N\Sigma^{\ast}}g^{(2)}_{\gamma\Sigma^{\ast0}\Sigma^0}$ & $g^{(1)}_{K^\ast\Delta\Sigma}$ \tabularnewline
\hline 
$-2.46(1.84)$ &$-2.34(1.98)$ &$157.21(167.35)$ &$92.39(63.02)$ &$4.10(6.38)$ &$25.23(17.61)$ & $40.00(4.83)$ \\
\hline
\end{tabular} 
\newline
\vspace*{0.2 cm}
\newline
\begin{tabular}{|c|c|c|c|c|c|c|}
\hline 
  &$M_R$ (MeV) & $\Gamma_R$ (MeV) & $f_1g_1$ & $f_2/f_1$ & $g_2/g_1$ & $g_3/g_1$ \tabularnewline
\hline 
$N^\ast(3/2^-)$& $2070.3(3.3)$& $78.5(9.1)$ & $-4.16(0.11)$ & $0.45(0.07)$ & $-0.45(0.01)$ & $1.43(0.01)$ \\
\hline
$N^\ast(7/2^-)$& $2433.3(17.6)$& $173.6(6.8)$ & $7.93(2.03)$ & $-2.07(0.35)$ & $-0.84(0.25)$ & $1.64(0.98)$ 
\\
\hline
\end{tabular}
\caption{Fitted parameters when adding one $N^\ast{3/2}^-$ and one $N^\ast{7/2}^-$ with $\chi^2/\text{ndf} = 1.10$.}\label{tab:solution_2rN3mN7m}
\end{table}

\begin{table}[h]
\begin{tabular}{|c|c|c|c|c|c|c|}
\hline 
  $g_{K^{\ast}N\Sigma}$ & $\kappa_{K^{\ast}N\Sigma}$ & $g^{(1)}_{K^{\ast}N\Sigma^{\ast}}g^{(1)}_{\gamma\Sigma^{\ast+}\Sigma^+}$ & $g^{(1)}_{K^{\ast}N\Sigma^{\ast}}g^{(2)}_{\gamma\Sigma^{\ast+}\Sigma^+}$ & $g^{(1)}_{K^{\ast}N\Sigma^{\ast}}g^{(1)}_{\gamma\Sigma^{\ast0}\Sigma^0}$ & $g^{(1)}_{K^{\ast}N\Sigma^{\ast}}g^{(2)}_{\gamma\Sigma^{\ast0}\Sigma^0}$ & $g^{(1)}_{K^\ast\Delta\Sigma}$ \tabularnewline
\hline 
$-2.46(1.82)$ &$-2.33(2.39)$ &$159.08(148.35)$ &$77.93(55.21)$ &$4.24(4.46)$ &$17.71(23.28)$ & $40.00(4.16)$ \\
\hline
\end{tabular} 
\newline
\vspace*{0.2 cm}
\newline
\begin{tabular}{|c|c|c|c|c|c|c|}
\hline 
  &$M_R$ (MeV) & $\Gamma_R$ (MeV) & $f_1g_1$ & $f_2/f_1$ & $g_2/g_1$ & $g_3/g_1$ \tabularnewline
\hline 
$N^\ast(3/2^-)$& $2070.6(3.1)$& $78.8(9.1)$ & $-4.12(0.07)$ & $0.47(0.04)$ & $-0.447(0.003)$ & $1.44(0.01)$ \\
\hline
$\Delta^\ast(7/2^-)$& $2437.1(14.0)$& $149.6(40.3)$ & $11.95(1.70)$ & $-2.08(0.25)$ & $-0.86(0.11)$ & $1.54(0.14)$ \\
\hline
\end{tabular}
\caption{Fitted parameters when adding one $N^\ast{3/2}^-$ and one $\Delta^\ast{7/2}^-$ with $\chi^2/\text{ndf} = 1.09$.}\label{tab:solution_2rN3mD7m}
\end{table}

\begin{figure}[h]
    \centering
    \includegraphics[width = 0.4\textwidth]{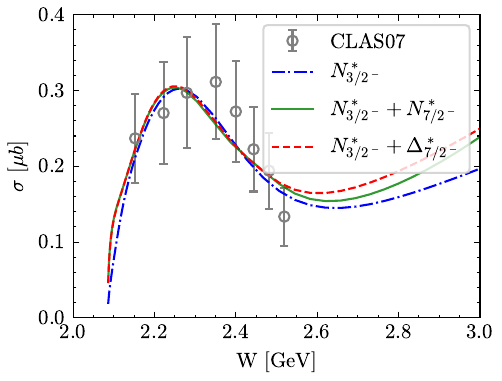}
    \hspace{0.8cm}
    \includegraphics[width = 0.4\textwidth]{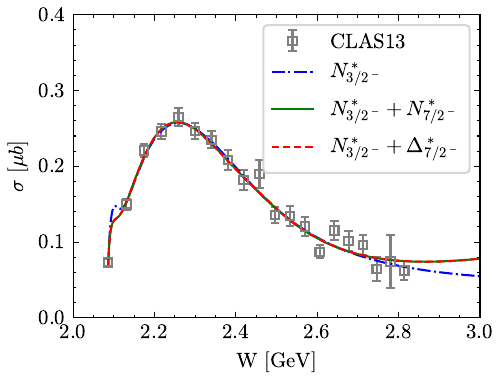}
    \caption{The total cross sections from the solution with adding one $N^\ast(3/2^-)$ (blue dash-dotted), one $N^\ast(3/2^-)$ plus one $N^\ast(7/2^-)$ (green solid), and one $N^\ast(3/2^-)$ plus one $\Delta^\ast(7/2^-)$ (red dashed), in comparison with the CLAS2007 $\gamma p\to K^{\ast 0}\Sigma^+$ (left) and CLAS2013 $\gamma p\to K^{\ast +}\Sigma^0$ (right) data, where $W$ indicates the c.m. energy.
    }
    \label{fig:tcs_ab}
\end{figure}

We show the solutions with two resonances in comparison with the experimental data in 
Fig.~\ref{fig:dcs_a} for $\gamma p\to K^{\ast 0}\Sigma^+$ reaction and Fig.~\ref{fig:dcs_b} for $\gamma p\to K^{\ast +}\Sigma^0$ reaction, where the solid and dashed lines indicate the solutions when adding one $N^\ast({3}/{2}^-)$ and one $N^\ast({7}/{2}^-)$, and one $N^\ast({3}/{2}^-)$ and one $\Delta^\ast({7}/{2}^-)$, respectively.
The total cross sections from these two solutions are shown in Fig.~\ref{fig:tcs_ab}, compared to the CLAS2007~\cite{CLAS:2007kab} $\gamma p\to K^{\ast 0}\Sigma^+$ and CLAS2013~\cite{CLAS:2013qgi} $\gamma p\to K^{\ast +}\Sigma^0$ reaction data.
From these figures, we can see that the experimental data can be well described by these solutions, although the improvement of adding the second resonance is relatively small.
We also see that the two solutions overlap to a large extent.
In principle, the combined fit of the $\gamma p\to K^{\ast 0}\Sigma^+$ and $\gamma p\to K^{\ast +}\Sigma^0$ reactions would be able to distinguish the nucleon and $\Delta$ resonances with the same quantum number, accounting the isospin factor that $I^N_{K^{\ast 0}N\Sigma^+}/I^N_{K^{\ast +}N\Sigma^0}=\sqrt{2}$ and $I^\Delta_{K^{\ast 0}N\Sigma^+}/I^\Delta_{K^{\ast +}N\Sigma^0}=1/\sqrt{2}$.
However, considering the two close solutions when adding two additional resonances,
it seems that the data are hard to distinguish the nucleon and $\Delta$ resonances with the same $J^P$.
This can be understandable considering the feature of the data we analyze, where we have 178 data points with small uncertainties for the $\gamma p\to K^{\ast +}\Sigma^0$ reaction, much more than the 72 data points for the $\gamma p\to K^{\ast 0}\Sigma^+$ reaction with relatively much larger errors.
Additionally, the energy region of the $\gamma p\to K^{\ast 0}\Sigma^+$ reaction concentrates upon $2.1$--$2.5$ GeV, while that for the $\gamma p\to K^{\ast +}\Sigma^0$ reaction lies in $2.1$--$2.8$ GeV.
From the total cross section plots of the $\gamma p\to K^{\ast 0}\Sigma^+$ reaction depicted in Fig.~\ref{fig:tcs_ab} on the left, we can see that the two solutions' descriptions are close in the analyzed region, while they separate in higher energies.
The discrepancies in the statistics, precision and energy region of the two datasets provide the possibility to balance the two solutions by adjusting the background and the common added $N^\ast(3/2^-)$ parameters.
The relatively small improvement in the description of the experimental data with the second $7/2^-$ resonance compared to adding only one $N^\ast(3/2^-)$ may be another reason for the two close results.
This problem can be resolved once the available data for the reactions $\gamma p\to K^{\ast +}\Sigma^0$ and $\gamma p\to K^{\ast 0}\Sigma^+$ have comparable statistics and precision, as well as a wider energy region.

Considering the facts that the best solution of adding one resonance needs a $N^\ast(3/2^-)$ resonance near $2097$ MeV and both of the best two solutions in the case of adding two resonances need a $N^\ast({3}/{2}^-)$ around $2070$ MeV,
it demonstrates the significant role of $N^\ast(2080)3/2^-$ in the $\gamma p\to K^\ast \Sigma$ reactions and supports the picture of the molecular bound state of the $K^\ast\Sigma$ system.
Reference~\cite{Ben:2023uev} considered the contributions of $N(2080)3/2^-$ and $N^\ast(2270)3/2^-$ as the S-wave $K^\ast\Sigma$ and $K^\ast\Sigma^\ast$ molecular states, respectively, and found that the data~\cite{CLAS:2007kab,CLAS:2013qgi} can be well described with a $1.648$ $\chi^2$ per data point.
The main difference of treating $N(2080)3/2^-$ is that, for the effective Lagrangian of the 
nucleon resonance with $K^\ast\Sigma$, Ref.~\cite{Ben:2023uev} used the pure $S$-wave form
\begin{equation}
    \mathcal{L}^{3/2^-}_{K^\ast\Sigma R}=g_{K^\ast\Sigma R}\bar{R}_\mu\Sigma K^{\ast\mu} + \mathrm{H.c.},
\end{equation}
which is equivalent to the $g_3$ term of Eq.~(\ref{eq:NsKsR3}) for $J^P=3/2^-$, while the terms with $g_1$ and $g_2$ therein incorporate higher partial-wave, i.e., $D$-wave, contributions.
Our better fitting result may suggest that higher partial-wave contributions are not negligible.
We give a summary table of the supported resonances from our primary solutions with their matching to the molecular model or possible PDG~\cite{ParticleDataGroup:2024cfk} listed particles in Table~\ref{tab:sum_match}, where the parameters of $N^\ast(3/2^-)$ are averaged from the three solutions.

\begin{table}[h]
    \centering
    \begin{tabular}{|c|c|c|c|
    }
    \hline
     & $M_R$ (MeV) & $\Gamma_R$ (MeV) & Matching\\
    \hline
        $N^\ast(3/2^-)$ & 2078.9(2.4) &74.5(5.4) & Molecular state of $K^\ast\Sigma$, $N(2080)$\\ 
    \hline
        $N^\ast(7/2^-)$ & 2433.3(17.6) & 173.6(6.8) &PDG, four-star $N(2190)$ \\
    \hline
        $\Delta^\ast(7/2^-)$ & 2437.1(14.0) & 149.6(40.3) & PDG, three-star $N(2200)$\\
    \hline
    \end{tabular}
    \caption{The supported resonances' parameters from our three solutions and their matching to the molecular state or PDG~\cite{ParticleDataGroup:2024cfk} listings.}
    \label{tab:sum_match}
\end{table}

\begin{figure}[h]
    \centering
    \includegraphics[width = 0.8\textwidth]{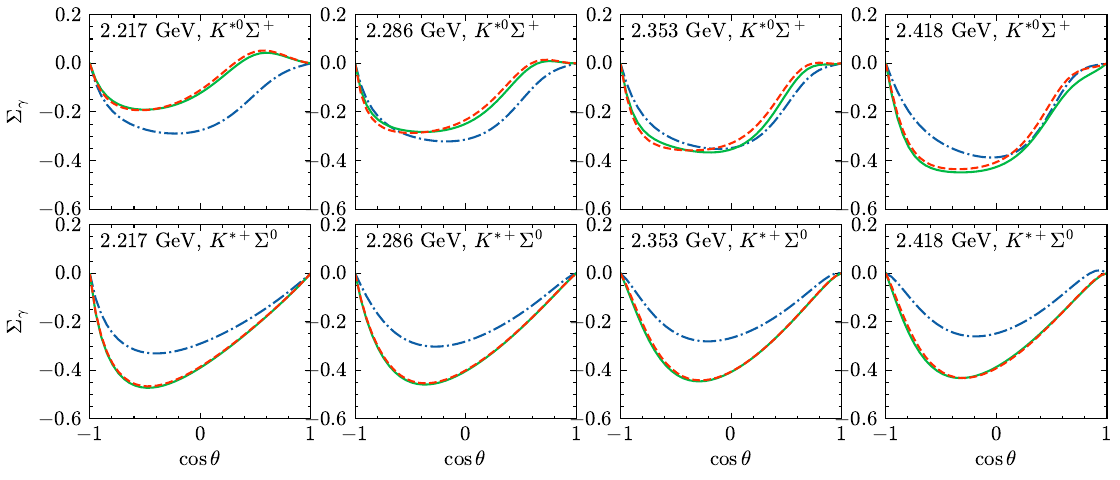}
    \caption{The predicted photon-beam asymmetry $\Sigma_\gamma$ for $\gamma p\to K^{\ast 0}\Sigma^+$ reaction (upper row) and $\gamma p\to K^{\ast +}\Sigma^0$ reaction (lower row) from the three solutions, when adding one $N(3/2^-)$ resonance (blue dash-dotted), one $N(3/2^-)$ resonance plus one $N(7/2^-)$ resonances (green solid), and one $N(3/2^-)$ resonance plus one $\Delta(7/2^-)$ resonance (red dashed).
    }
    \label{fig:pol_S_g}
\end{figure}

\begin{figure}[h]
    \centering
    \includegraphics[width = 0.8\textwidth]{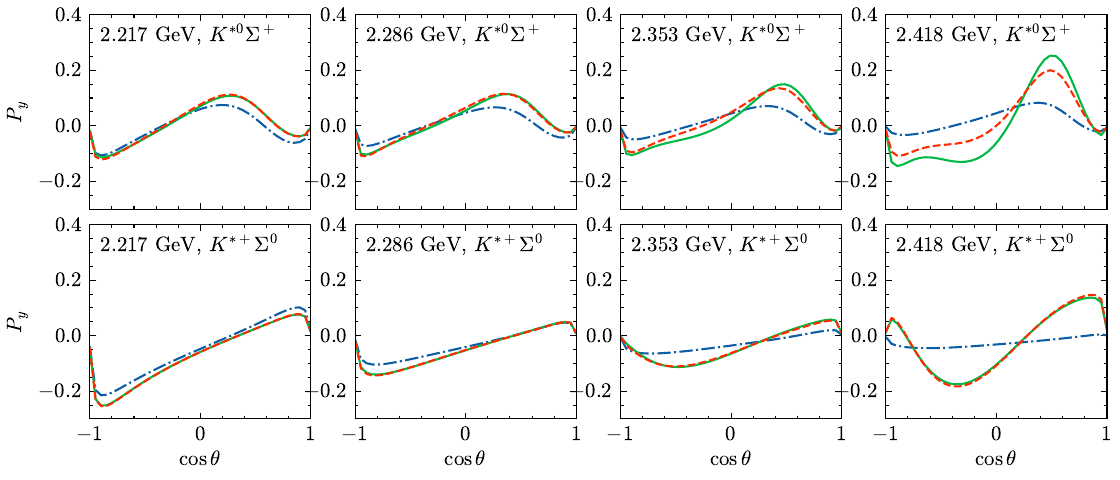}
    \caption{The predicted recoil asymmetry $P_y$ for $\gamma p\to K^{\ast 0}\Sigma^+$ reaction (upper row) and $\gamma p\to K^{\ast +}\Sigma^0$ reaction (lower row) from the three solutions with the same notation as Fig.~\ref{fig:pol_S_g}.
    }
    \label{fig:pol_P_y}
\end{figure}

\begin{figure}[h]
    \centering
    \includegraphics[width = 0.8\textwidth]{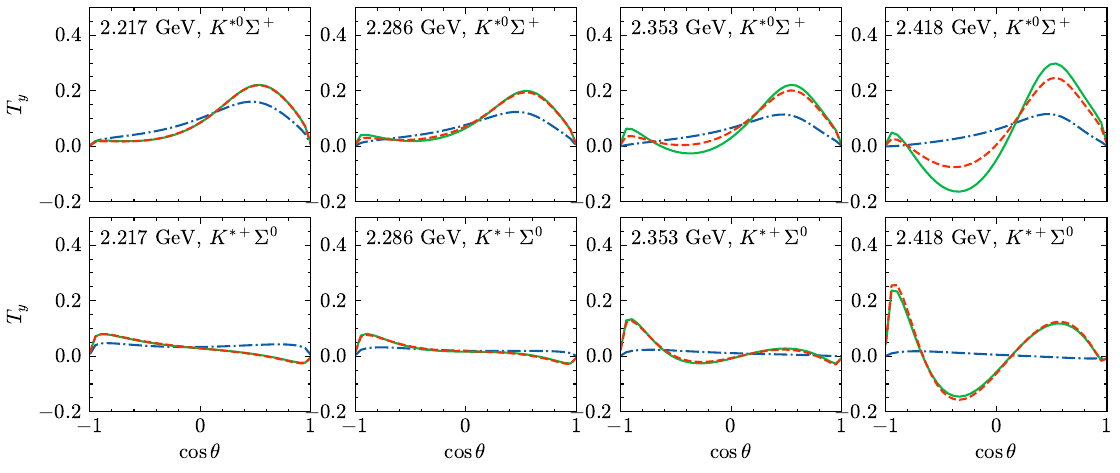}
    \caption{The predicted target asymmetry $T_y$ for $\gamma p\to K^{\ast 0}\Sigma^+$ reaction (upper row) and $\gamma p\to K^{\ast +}\Sigma^0$ reaction (lower row) from the three solutions noted the same as Fig.~\ref{fig:pol_S_g}.
    }
    \label{fig:pol_T_y}
\end{figure}

Currently, polarization measurements of the $\gamma p\to K^{\ast}\Sigma$ reactions have not been reported.
We can use our solutions to estimate the behavior of the polarization observables.
The single polarization observables of $\gamma p\to K^{\ast}\Sigma$ include the photon-beam asymmetry $\Sigma_\gamma$, recoil asymmetry $P_y$, and target asymmetry $T_y$ which can be expressed as~\cite{Pichowsky:1994gh,Titov:2002iv}
\begin{eqnarray}
    \Sigma_\gamma &\equiv& \frac{d\sigma(\epsilon_{\perp})-d\sigma(\epsilon_{\parallel})}{d\sigma_{\text{unpol}}},\\
    P_y &\equiv&\frac{d\sigma(s^\Sigma_y=\frac{1}{2})-d\sigma(s^\Sigma_y=-\frac{1}{2})}{d\sigma_{\text{unpol}}},\\
    T_y &\equiv&\frac{d\sigma(s^N_y=\frac{1}{2})-d\sigma(s^N_y=-\frac{1}{2})}{d\sigma_{\text{unpol}}}, 
\end{eqnarray}
where $d\sigma_{\text{unpol}}$ indicates the unpolarized differential cross section, and $s^B_y=1/2 (-1/2)$ indicates the spin of the baryon $B$ along the $y$ axis, which is perpendicular to the reaction plane, in which the positive $z$ axis is along the photon three-momentum direction in the c.m. frame.
The predicted polarization values of $\Sigma_\gamma$, $P_y$, and $T_y$ are plotted in the upper and lower panels for the $\gamma p\to K^{\ast 0}\Sigma^+$ and $\gamma p\to K^{\ast +}\Sigma^0$ reactions in Figs.~\ref{fig:pol_S_g}--\ref{fig:pol_T_y}, respectively, where the blue dash-dotted lines, the solid green lines, and the red dashed lines represent the solution of adding only $N^\ast(3/2)^-$, adding $N^\ast(3/2)^-$ and $N^\ast(7/2)^-$, and adding $N^\ast(3/2)^-$ and $\Delta^\ast(7/2)^-$, separately.
We can see that the solution of adding only $N^\ast(3/2)^-$ differs from other solutions, while the two solutions of adding two resonances overlaps a lot except at the c.m. energy of $2.35$ and $2.42$ GeV.
Thus, the experimental polarization data including the c.m. energy at $2.42$ GeV can help to distinguish the importance of the second added $N^\ast(7/2)^-$ or $\Delta^\ast(7/2)^-$ resonance.
From a rough estimation, our results suggest that, with an experimental precision of 10\%$--$20\%, the two solutions can be resolved at a statistical significance of 3 standard deviations ($3\sigma$) around $\cos{\theta}= 0.5$ and $\cos{\theta}= -0.3$ at $2.42$ GeV in the $P_y$ and $T_y$ polarizations.

\section{Summary and Outlook}\label{sec:summary}
We have analyzed all the available differential cross section data of $\gamma p\to K^{\ast 0}\Sigma^+$ and $\gamma p\to K^{\ast +}\Sigma^0$ reactions carrying out a different strategy from previous literature, that is instead of adding the additional resonance with fixed parameters either from PDG or models, we investigate the additional resonance with specific $J^P$ while keeping its parameters to be determined by experimental data.
The background contribution includes the $t$-channel $\kappa$, $K$, and $K^\ast$ exchanges, the $u$-channel $\Lambda$, $\Sigma$, and $\Sigma(1385)$ exchanges, and the $s$-channel nucleon and $\Delta(1232)1/2^+$ exchanges.
With only the background contributions, the fitted result is $\chi^2/\text{ndf}=3.40$ and cannot adequately describe the experimental data.
When only adding one additional resonance, the best solution is to add one $N^\ast(3/2)^-$ around $2.097$ GeV, which substantially reduces the $\chi^2/\text{ndf}$ to $1.35$.
When adding two additional resonances, one of the best two solutions is to add one $N^\ast(3/2)^-$ and one $N^\ast(7/2)^-$, the other is to add one $N^\ast(3/2)^-$ and one $\Delta^\ast(7/2)^-$, with $\chi^2/\text{ndf}$ being $1.10$ and $1.09$, respectively.
Both solutions need a $N^\ast(3/2)^-$ resonance around $2070$ MeV.
Associated with the best solution in the case of adding one resonance where a $N^\ast(3/2)^-$ resonance at $2097$ MeV is needed, these results suggest that the $N(3/2)^-$ resonance near $2080$ MeV provides significant contributions to the $\gamma p\to K^\ast\Sigma$ reactions and is strongly coupled with the $K^\ast\Sigma$ final states.
Our results support the possibility of $N(2080)3/2^-$ as the molecular bound state of the $K^\ast\Sigma$ system.
According to the solutions when adding two additional resonances which need one $N^\ast(3/2^-)$ resonance and one $7/2^-$ nucleon or $\Delta$ resonance near $2433$ MeV,
it seems that the experimental data used here are difficult to distinguish the two $7/2^-$ resonances.
This condition can be improved when the differential cross section data of the $\gamma p\to K^{\ast 0}\Sigma^+$ and $\gamma p\to K^{\ast +}\Sigma^0$ reactions have comparable statistics, precision, and energy regions and when the polarization data are available.
Although a dynamical coupled-channel analysis that retains unitarity would be more proper,
it needs a simultaneous fit to all the involved channels, which is not realistic at present due to the lack of experimental data at such higher energies.
Since the resonances incorporated are not broad, the procedure we carry out is a reasonable approximation, and
we expect the influence from the couple-channel analysis on the resonance parameters to be very small.
We show the predictions of the polarization observables with our three solutions in Figs.~\ref{fig:pol_S_g}--\ref{fig:pol_T_y}.
We hope future experiments with higher statistic and polarization observables can help to distinguish the role of the nucleon and $\Delta$ resonances and to further verify the molecular identity of $N^\ast(2080)3/2^-$.

\begin{acknowledgments}
The authors thank Jian Liang, Ai-Chao Wang, Di Ben, Fei Huang, Yu Lu, Jia-Jun Wu, Feng-Kun Guo and Qian Wang for helpful discussions.
This work is supported by the Natural Science Foundation of China under Grant No. 12105108 and the Guangdong Major Project for Basic and Applied Basic Research under Grant No. 2020B0301030008.
\end{acknowledgments}

\bibliography{library}

\end{document}